\documentclass[twocolumn]{jpsj2}
\usepackage{graphicx}
\usepackage{amsmath,amssymb}

\def\runauthor{Youichi {\sc Yanase}, M. Motizuki and Masao {\sc Ogata}}

\title{ 
Multi-orbital Analysis on the Superconductivity in 
${\rm Na_{x}Co_{}O_{2}} \cdot y{\rm H}_{2}{\rm O}$
} 

\author{Youichi {\sc Yanase}\footnote{E-mail:
yanase@hosi.phys.s.u-tokyo.ac.jp}, Masahito {\sc Mochizuki} and Masao {\sc Ogata}}

\inst{Department of Physics, University of Tokyo, Tokyo 113-0033}

\recdate{15 July 2004}

\abst
{
 We preform a multi-orbital analysis on the novel superconductivity
in ${\rm Na_{x}Co_{}O_{2}} \cdot y{\rm H}_{2}{\rm O}$. 
 We construct a three-orbital model which reproduces the band structure 
expected from the LDA calculation. 
 The effective interaction leading to the pairing is estimated by means
of the perturbation theory. 
 It is shown that the spin triplet superconductivity is stabilized 
in the wide parameter region. This is basically owing to the 
ferromagnetic character of spin fluctuation. 
 The $p$-wave and $f$-wave superconductivity are nearly degenerate. 
 The former is realized when the Hund's rule coupling is large, and 
vice versa. 
 In a small part of the parameter space, the $d$-wave superconductivity 
is also stabilized. 
 We point out that the orbital degeneracy plays an essential role 
for these results. In particular, the momentum dependence of 
wave function of quasi-particles is quite important. 
 The nearly degeneracy of $p$-wave and $f$-wave superconductivity 
is explained by analysing the microscopic character of each Fermi surface.  
 We discuss the validity of some reduced models. 
 While the single-orbital Hubbard model reproducing  
the Fermi surface is qualitatively inappropriate,  we find an 
effective two-orbital model appropriate for studying the 
superconductivity. 
 We investigate the vertex corrections higher than the third order on 
the basis of the two-orbital model. 
 It is shown that the vertex correction induces the screening effect 
but does not affect qualitative results. 
}

\kword
{
${\rm Na_{x}Co_{}O_{2}} \cdot y{\rm H}_{2}{\rm O}$; 
unconventional superconductivity; 
ferromagnetic spin fluctuation; multi-orbital analysis
}

\begin{document}
\sloppy
\maketitle

\newcommand{\eli}{$\acute{{\rm E}}$liashberg }
\renewcommand{\k}{\mbox{\boldmath$k$}}
\newcommand{\q}{\mbox{\boldmath$q$}}
\newcommand{\Q}{\mbox{\boldmath$Q$}}
\newcommand{\kk}{\mbox{\boldmath$k'$}}
\newcommand{\e}{\varepsilon}
\newcommand{\ee}{\varepsilon^{'}}
\newcommand{\s}{{\mit{\it \Sigma}}}
\newcommand{\J}{\mbox{\boldmath$J$}}
\newcommand{\vv}{\mbox{\boldmath$v$}}
\newcommand{\Jh}{J_{{\rm H}}}
\newcommand{\LL}{\mbox{\boldmath$L$}}
\renewcommand{\SS}{\mbox{\boldmath$S$}}
\newcommand{\Tc}{$T_{\rm c}$ }
\newcommand{\Tcf}{$T_{\rm c}$}
\newcommand{\Co}{${\rm Na_{x}Co_{}O_{2}} \cdot y{\rm H}_{2}{\rm O}$ }
\newcommand{\Cof}{${\rm Na_{x}Co_{}O_{2}} \cdot y{\rm H}_{2}{\rm O}$}
\newcommand{\tgf}{$t_{\rm 2g}$-orbitals}
\newcommand{\tg}{$t_{\rm 2g}$-orbitals }
\newcommand{\av}{\mbox{\boldmath${\rm a}$} }
\newcommand{\bv}{\mbox{\boldmath${\rm b}$} }
\newcommand{\avf}{\mbox{\boldmath${\rm a}$}}
\newcommand{\bvf}{\mbox{\boldmath${\rm b}$}}
\newcommand{\egf}{$e_{\rm g}$-Fermi surface }
\newcommand{\egff}{$e_{\rm g}$-Fermi surface}
\newcommand{\agf}{$a_{\rm 1g}$-Fermi surface }
\newcommand{\agff}{$a_{\rm 1g}$-Fermi surface}

\section{Introduction}

 Since the discovery of High-\Tc superconductivity~\cite{rf:bednortz} 
and heavy fermion superconductors~\cite{rf:steglich}, 
the mechanism of superconductivity induced by electron correlation 
has been one of the central issues in the condensed matter physics. 
 In this study, recently discovered superconductor \Co 
is analyzed in details.

 Immediately after the discovery of superconductivity 
in water-intercalated Cobalt oxides 
${\rm Na_{x}Co_{}O_{2}} \cdot y{\rm H}_{2}{\rm O}$,~\cite{rf:takada} 
both experimental~\cite{rf:ong,rf:sugiyama,
rf:chou,rf:yoshimura,rf:kobayashi,rf:zheng,rf:ishida,rf:higemoto,
rf:motohashi,rf:li,rf:miyosi,rf:uemura,rf:kanigel,rf:hdyang,rf:lorenz,
rf:oeschler,rf:sakurai} and 
theoretical~\cite{rf:Atanaka,rf:koshibae,rf:baskaran,rf:shastry,rf:lee,
rf:ogata,rf:ikeda,rf:Ytanaka,rf:honerkamp,rf:kuroki,rf:nisikawa,rf:motrunich} 
studies have been performed extensively. 
 While some controversial results exist, many experimental evidences 
for the non-$s$-wave superconductivity~\cite{rf:sakurai} 
has been reported by NMR~\cite{rf:yoshimura,rf:kobayashi,
rf:zheng,rf:ishida} and specific heat measurements.~\cite{rf:hdyang,
rf:lorenz,rf:oeschler} 
 The characteristic behaviors in strongly correlated electron 
systems have been observed in the non-water-intercalated 
compounds.~\cite{rf:chou,rf:li,rf:miyosi,rf:ando}
 The existence of the magnetic 
phase~\cite{rf:motohashi,rf:ong,rf:sugiyama} in ${\rm Na_{x}Co_{}O_{2}}$ 
with $x \sim 0.75$ also indicates an importance of electron correlation. 
 These compounds have a layered structure like cuprate~\cite{rf:bednortz} 
and ruthenate~\cite{rf:maeno}, and the two-dimensionality is enhanced by
the water-intercalation. 
 These circumstantial evidences indicate that \Co is an 
unconventional superconductor induced by the electron correlation.

 The theoretical interests are turned on also by the symmetry of 
crystal structure. 
 In contrast to the square lattice in cuprates and ruthenates, 
the layer is constructed from the triangular lattice of Co ions. 
 Then, a novel symmetry of Cooper pairing is possible in principle. 
 The $d$-wave superconductivity in cuprate superconductors and 
$p$-wave superconductivity in ruthenates have been established before. 
 In addition to them, the spin triplet $f$-wave superconductivity 
and spin singlet $i$-wave one are possible from the analysis of 
pairing symmetry (see Table.~I).

 The effect of frustration, which is characteristic in the spin system 
on the triangular lattice, has also attracted much attention. 
 The RVB theory has been applied to the triangular 
lattice~\cite{rf:baskaran,rf:shastry,rf:lee,rf:ogata} and 
basically concluded the spin singlet $d$-wave superconductivity. 
 Then, $d_{\rm x^{2}-y^{2}} \pm$ i$d_{\rm xy}$-wave symmetry is 
expected below \Tc owing to the six-fold symmetry of triangular lattice. 
 However, the time-reversal symmetry breaking has not been observed 
until now.~\cite{rf:higemoto} 
 Some authors have pointed out the frustration of charge ordering 
for the electron filling $n=4/3$,~\cite{rf:lee} and 
the $f$-wave superconductivity due to the charge fluctuation 
has been discussed.~\cite{rf:Ytanaka,rf:motrunich}

 Another interesting property of \Co is the orbital degeneracy. 
 The conduction band of this material mainly consists of 
three $t_{\rm 2g}$-orbitals in Co ions which hybridize with 
O2p-orbitals. 
 Thus far, most of theoretical studies on the superconductivity 
have been performed on the basis of the single-orbital model. 
 These investigations have successfully achieved microscopic 
understandings on the cuprate, organic and ruthenate 
superconductors.~\cite{rf:yanasereview} 
 However, we consider that the theoretical analysis including the orbital 
degeneracy is highly desired in order to understand a variety of 
superconductors including \Co and 
heavy fermion compounds. 
 The superconductivity in $d$-electron systems provides a favorable 
subject for the theoretical development along this line, 
because a simple electronic structure is expected compared to 
heavy fermion superconductors. 
 Although Sr$_2$RuO$_4$ has been a precious compound in this sense,  
then the orbital degree of freedom is not important for the basic 
mechanism of superconductivity.~\cite{rf:nomura,rf:yanaseRuSO} 
 In this study, we show that the orbital degeneracy plays 
an essential role in \Co in contrast to the 
ruthenate superconductor. 
 We conclude that \Co is a typical multi-orbital superconductor 
in this sense.

 We adopt a perturbative method for the unconventional 
superconductivity,~\cite{rf:yanasereview} which is a systematic 
approach for the electron correlation. 
 Note that the spin fluctuation theory~\cite{rf:moriyaAD} which 
is widely used for superconductivity is microscopically formulated 
in this method. 
 It is expected that this approach is reliable from weak to 
intermediate coupling region. 
 Before the discovery of \Cof, this method has been applied to the 
single-orbital triangular lattice model. 
 Then, the $d$-wave,~\cite{rf:vojta} $f$-wave~\cite{rf:kuroki2001} and 
$p$-wave superconductivity~\cite{rf:nisikawa2002} have been obtained. 
 Some authors have applied this calculation to \Cof, and reported 
the spin singlet $d$- or $i$-wave superconductivity,~\cite{rf:honerkamp} 
spin triplet $f$-wave superconductivity~\cite{rf:kuroki} 
and nearly degeneracy between $d$- and $f$-wave 
superconductivity.~\cite{rf:nisikawa} 
 We consider that this puzzling problem should be resolved 
by the multi-orbital analysis involving the microscopic aspects of 
electronic structure.

 In this paper, we analyze a multi-orbital Hubbard model constructed from  
three Co $t_{\rm 2g}$-orbitals. 
 This model appropriately reproduces the electronic structure obtained 
in the LDA calculation.~\cite{rf:singh,rf:pickett} 
 The wave function of quasi-particles, which is neglected in the 
single-orbital Hubbard models, is appropriately taken into account 
in this multi-orbital model. We show that the momentum dependence 
of this wave function plays an essential role for the mechanism of 
superconductivity. 
 We determine the most stable superconducting state with use of 
the perturbation theory. 
 According to the results of second order perturbation (SOP), 
third order perturbation (TOP) and renormalized third order perturbation 
(RTOP) theories, it is concluded that the spin triplet $p$-wave or $f$-wave 
superconductivity is stable in the wide region of parameter space. 
 The pairing interaction is closely related to the ferromagnetic 
character of spin susceptibility, although the pairing interaction is 
not simply described by the spin susceptibility like 
in the single-orbital model.~\cite{rf:yanasereview} 
 While the momentum dependence of spin susceptibility is usually not 
remarkable in the frustrating system, the ferromagnetic character 
clearly appears in the present case owing to the orbital degree of freedom.

 From a comparison with single-orbital Hubbard models, 
the important roles of orbital degeneracy are illuminated in \S4.1. 
 Alternatively, we propose a reduced two-orbital model including the 
$e_{\rm g}$-doublet in \S4.2. 
 It is shown that results for the superconductivity is appropriately 
reproduced in this simplified model. 
 On the basis of the two-orbital model, we investigate the roles of  
vertex correction terms in \S5. 
 Then, we show that the vertex correction term, which significantly 
enhances the spin triplet pairing in Sr$_{2}$RuO$_{4}$,~\cite{rf:nomura} 
is not important in case of \Cof. 
 Thus, the superconducting instability is basically described within 
the SOP. 
 Therefore, we first explain in details the results of SOP in \S3, 
and discuss the reduced models in \S4 and the role of vertex 
corrections in \S5.

\section{Multi-orbital model}

 First, we construct a multi-orbital model for \Cof.~\cite{rf:motiduki} 
 We consider a two-dimensional model which represents the Co ions on the 
triangular lattice. 
 Note that the superconductivity occurs when the two-dimensionality 
is enhanced by the water-intercalation. 
 We also note that the conduction band mainly consists of Co 
\tgf.~\cite{rf:singh,rf:pickett} 
 Co ion is enclosed by an octahedron of oxygens and 
nearest neighbor Co ions share the edge of the octahedron. 
 We describe the dispersion relation by using a tight-binding model 
and adopt a multi-orbital Hubbard Hamiltonian written as, 
\begin{eqnarray}
&& H_{3} = H_{0}+H_{{\rm I}}, 
\\
&& H_{0} = \sum_{i,j,s} \sum_{a,b} t_{a,b,i,j} 
c_{i,a,s}^{\dag} c_{j,b,s},  
\\
&& H_{{\rm I}} = 
U \sum_{i} \sum_{a} n_{i,a,\uparrow} n_{i,a,\downarrow} 
+ U' \sum_{i} \sum_{a>b} n_{i,a} n_{i,b} 
\nonumber \\
&& \hspace{10mm}
- \Jh \sum_{i} \sum_{a>b} (2 \SS_{i,a} \SS_{i,b} + \frac{1}{2} n_{i,a} n_{i,b})
\nonumber \\
&& \hspace{10mm}
+ J \sum_{i} \sum_{a \neq b}  
c_{i,a,\downarrow}^{\dag} 
c_{i,a,\uparrow}^{\dag} 
c_{i,b,\uparrow} 
c_{i,b,\downarrow}. 
\label{eq:multi-orbital-model}
\end{eqnarray}

 The first term $H_{0}$ is a tight-binding Hamiltonian where 
$t_{a,b,i,j}$ are hopping matrix elements. 
 Here, the indices $i$ and $j$ denote the sites in the real space and 
indices $a$ and $b$ denote the orbitals. 
 We assign the $d_{\rm xy}$-, $d_{\rm yz}$- and $d_{\rm xz}$-orbitals 
to $a=1$, $a=2$ and $a=3$, respectively. 
 The largest matrix element is the inter-orbital hopping through 
O2p-orbitals, which are $t_{1,2,i,j}$ for $j=i \pm$(\avf+\bvf), 
$t_{2,3,i,j}$ for $j=i \pm$\av and $t_{1,3,i,j}$ for $j=i \pm$\bvf. 
 We choose the lattice constant as a unit length and denote 
the unit vectors as \avf=$(\sqrt3/2,-1/2)$ and \bvf=$(0,1)$ which 
are the basis of the triangular lattice. 
 If we assume only the largest matrix elements, the system is regarded 
to be a superposition of the kagome lattice.~\cite{rf:koshibae} 
 However, the long range hopping through the O2p-orbitals and 
direct hopping between Co ions are necessary to reproduce the Fermi 
surface obtained in the LDA calculation.

 We take account of the matrix elements within third-nearest-neighbor 
sites according to the symmetry of orbitals and lattice. 
 They are described by nine parameters from $t_1$ to $t_9$. 
 The non-interacting Hamiltonian is described in the matrix
representation,  
\begin{eqnarray}
 \label{eq:three-band-model-kinetic}
&& H_0 = \sum_{\k,s} c_{\k,s}^{\dag} \hat{H}(\k) c_{\k,s}, 
\\
&& \hat{H}(\k) =
\left(
\begin{array}{ccc}
\varepsilon_{11}(\k) & \varepsilon_{12}(\k) & \varepsilon_{13}(\k)\\
\varepsilon_{21}(\k) & \varepsilon_{22}(\k) & \varepsilon_{23}(\k)\\
\varepsilon_{31}(\k) & \varepsilon_{32}(\k) & \varepsilon_{33}(\k)\\
\end{array}
\right),  
\end{eqnarray}
 where $c_{\k,s}^{\dag}=
(c_{\k,1,s}^{\dag},c_{\k,2,s}^{\dag},c_{\k,3,s}^{\dag})$ is a vector 
representation of the Fourier transformed creation operators with spin $s$. 
 The matrix elements are obtained as, 
\begin{eqnarray}
\label{eq:e11}
&& \hspace{-10mm}
\varepsilon_{11}(\k) = 2 t_1 \cos k_1 + 2 t_2 (\cos k_2 +\cos k_3) 
\nonumber \\
&& \hspace{-5mm}
+2 t_4 (\cos(k_1 - k_3)+\cos(k_1-k_2)) + 2 t_5 \cos 2 k_1, 
\\
&& \hspace{-10mm}
\varepsilon_{22}(\k) = 2 t_1 \cos k_2 + 2 t_2 (\cos k_1 +\cos k_3) 
\nonumber \\ 
&& \hspace{-5mm}
+2 t_4 (\cos(k_1 - k_2)+\cos(k_2 - k_3)) + 2 t_5 \cos 2 k_2, 
\\
&& \hspace{-10mm}
\varepsilon_{33}(\k) = 2 t_1 \cos k_3 + 2 t_2 (\cos k_1 +\cos k_2) 
\nonumber \\ 
&& \hspace{-5mm}
+2 t_4 (\cos(k_1 - k_3)+\cos(k_2 - k_3)) + 2 t_5 \cos 2 k_3, 
\\
&& \hspace{-10mm}
\varepsilon_{12}(\k) = 2 t_3 \cos k_3 + 2 t_6 \cos 2 k_3  
+2 t_7 \cos(k_1 - k_3) 
\nonumber \\ 
&& \hspace{0mm} 
+ 2 t_8 \cos(k_2 - k_3) + t_9 \cos(k_1-k_2)
- e_{\rm c}/3, 
\\
&& \hspace{-10mm}
\varepsilon_{13}(\k) = 2 t_3 \cos k_2 + 2 t_6 \cos 2 k_2  
+2 t_7 \cos(k_2 - k_3)
\nonumber \\ 
&& \hspace{0mm}
+ 2 t_8 \cos(k_1 - k_2) + t_9 \cos(k_1 - k_3)
- e_{\rm c}/3, 
\\
&& \hspace{-10mm}
\varepsilon_{23}(\k) = 2 t_3 \cos k_1 + 2 t_6 \cos 2 k_1  
+2 t_7 \cos(k_1 - k_2)
\nonumber \\ 
\label{eq:e23}
&& \hspace{0mm}
+ 2 t_8 \cos(k_1 - k_3) + t_9 \cos(k_2 - k_3)
- e_{\rm c}/3, 
\end{eqnarray}
where $k_1=\sqrt{3}/2 k_{\rm x} - 1/2 k_{\rm y}$, $k_2=k_{\rm y}$ and 
$k_3=-k_1-k_2$. 
 The parameter $e_{\rm c}$ represents the crystal field splitting of 
\tg arising from the distortion of octahedron. 
 A typical dispersion relation and Fermi surface are shown in Fig.~1. 
 There is a hole pocket enclosing the $\Gamma$-point and six hole
pockets near the K-points, which are consistent with LDA 
calculations.~\cite{rf:singh,rf:pickett} 
 We choose the unit of energy as $t_{3}=1$ throughout this paper.

 Although $e_{\rm c}$ seems to be small, it is useful to use 
a non-degenerate $a_{\rm 1g}$-orbital and doubly-degenerate 
$e_{\rm g}$-orbitals. 
 They are defined from the three \tg as 
\begin{eqnarray} 
\label{eq:e1}
|e_{\rm g}, 1> = \frac{1}{\sqrt{2}}(|{\rm xz}>-|{\rm yz}>),
\\
\label{eq:e2}
|e_{\rm g}, 2> = \frac{1}{\sqrt{6}}(2|{\rm xy}>-|{\rm xz}>-|{\rm yz}>),
\\
\label{eq:a1g}
|a_{\rm 1g}> = \frac{1}{\sqrt{3}}(|{\rm xy}>+|{\rm xz}>+|{\rm yz}>).
\end{eqnarray}
 The wave function of $a_{\rm 1g}$-orbital spreads along the 
{\it c}-axis, and those of $e_{\rm g}$-orbitals spread along the 
two-dimensional plane. 
 We will show later that this representation is appropriate for 
understanding the mechanism of superconductivity (\S4.2).

\begin{figure}[ht]
\begin{center}
\includegraphics[width=7cm]{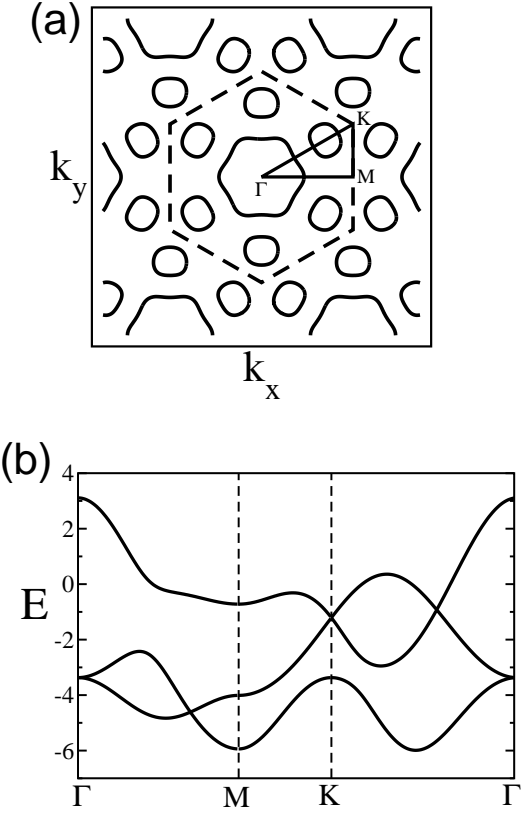}
\caption{(a) Fermi surfaces and (b) dispersion relation 
obtained from the tight-binding Hamiltonian. 
The dashed line in (a) shows the first Brillouin zone. 
The parameters are chosen to be 
$(t_1,t_2,t_3,t_4,t_5,t_6,t_7,t_8,t_9)=
(0.08,0.16,1,0.24,-0.16,-0.04,0.16,0.16,-0.2)$. 
} 
\label{fig:fermisurface}
\end{center}
\end{figure}

 The hole pocket around the $\Gamma$-point in 
Fig.~\ref{fig:fermisurface}(a) mainly consists of 
the $a_{\rm 1g}$-orbital and the six hole pockets near the K-points 
mainly consist of the $e_{\rm g}$-orbitals. 
 Thus, we denote these Fermi surfaces as 
$a_{\rm 1g}$-Fermi surface and $e_{\rm g}$-Fermi surface, respectively. 
 This nature of the Fermi surface is consistent with LDA 
calculations.~\cite{rf:singh,rf:pickett} 
 Note that recent ARPES measurements~\cite{rf:hasan,rf:yang} 
for non-superconducting Na$_x$CoO$_{2}$ observed the $a_{\rm 1g}$-Fermi 
surface, but the $e_{\rm g}$-Fermi surface has not been found. 
 Fermi surface of water-intercalated Na$_x$CoO$_{2}$ is not clear 
at present. 
 Moreover, the valence of Co ion in superconducting materials is also 
under debate.~\cite{rf:karppinen} 
 Therefore, we investigate a wide region in the parameter space and 
study the possible pairing instability. 
 It is one of the goals of this paper to study the relation between 
the electronic state and superconductivity. 
 It will be shown that the superconductivity is hard to be stabilized  
when $e_{\rm g}$-Fermi surface vanishes.

 The second term $H_{{\rm I}}$ describes the short range 
Coulomb interactions which include the intra-orbital repulsion $U$, 
inter-orbital repulsion $U'$, Hund's rule coupling $\Jh$ and pair hopping term 
$J$. 
 The relations $U=U'+\Jh+J$ and $\Jh=J$ are satisfied in a simple 
estimation. 
 Under these conditions, the interaction term $H_{{\rm I}}$ is 
invariant for the local unitary transformation between orbitals 
which will be used later. 
 If these relations are violated, the symmetry of triangular lattice is 
artificially broken.  
 Therefore, we impose these relations through this paper. 
 Although possible roles of the long range Coulomb interaction have been 
investigated,~\cite{rf:baskaran,rf:Ytanaka,rf:motrunich} we concentrate on 
the short range interaction in this paper.

 Note that previous studies based on a perturbative method 
for cuprates, organics and ruthenate have succeeded in identifying 
the dominant scattering process leading to the 
superconductivity.~\cite{rf:yanasereview} 
 This theory is complementary to the fluctuation 
theory which is represented by a random phase approximation (RPA) or 
fluctuation exchange approximation (FLEX). 
 Generally speaking, the fluctuation theory will be appropriate 
in the vicinity of the magnetic or other instabilities, 
because the critical enhancement of the fluctuation is taken into account. 
 On the other hand, the perturbation theory is more appropriate 
when the critical enhancement of any particular fluctuation is absent, 
because all terms in the same order are taken into account without any 
prejudice. 
 We perform the second order perturbation as well as 
the third order perturbation in this paper. 
 The results of FLEX study will be published 
elsewhere,~\cite{rf:motiduki} where qualitatively consistent results 
are obtained.

\section{Second Order Perturbation}

\subsection{Details of calculation and classification of pairing symmetry}

 In this section, we investigate the superconducting instability 
by using the \eli equation within the second order perturbation (SOP). 
 The basic procedure has been explained in 
literatures~\cite{rf:yanasereview} and the extension to 
multi-orbital model is straightforward. 
 The \eli equation is described by the Green function and 
the effective interaction. The latter is represented by an irreducible 
four point vertex in the particle-particle channel (Fig.~2(a)). 
 The second order terms in the effective interaction are 
diagrammatically represented by Figs.~2(b-e). 
 In case of the single-orbital Hubbard model, this term is simply 
expressed as 
$V(k,k')=U^{2}\chi_{0}(k-k')$ for spin singlet pairing and 
$V(k,k')=-U^{2}\chi_{0}(k-k')$ for spin triplet pairing, respectively, 
with a bare spin susceptibility $\chi_{0}(k-k')$.  
 However, in the multi-orbital model, the four point vertex has indices of 
orbitals as $V_{abcd}(k,k')$ (see Fig.~2(a)),  
which is calculated from the possible 
combination of Coulomb interactions and Green functions.

\begin{figure}[ht]
\begin{center}
\includegraphics[width=8cm]{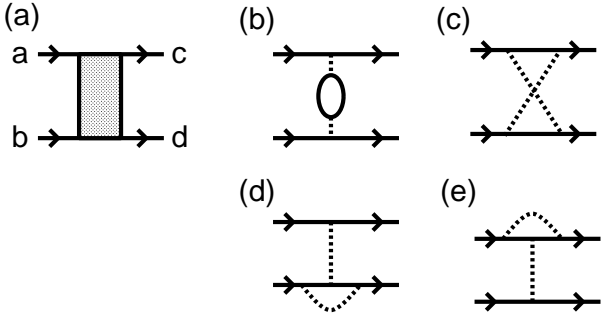}
\caption{
(a) Diagrammatic representation of the effective interaction leading 
to the superconductivity. 
(b-e) The second order terms with respect to the Coulomb interactions 
(dashed lines). The solid line denotes the Green function having the 
indices of spin and orbital. 
} 
\label{fig:diagram}
\end{center}
\end{figure}

 In order to make the following discussions clear, we introduce a 
unitary transformation $\hat{U}(\k)=(u_{ij}(\k))$ which 
diagonalizes $\hat{H}(\k)$, namely 
\begin{eqnarray}
 \label{eq:unitary}
\hat{U}^{\dag}(\k)  \hat{H}(\k) \hat{U}(\k) 
= 
\left(
\begin{array}{ccc}
E_1(\k) & 0 & 0\\
0 & E_2(\k) & 0\\
0 & 0 & E_3(\k)\\
\end{array}
\right).
\end{eqnarray}
 Here, we choose $E_1(\k) \leq E_2(\k) \leq E_3(\k)$. 
 With use of these matrix elements, the matrix form of 
Green function characterized by orbitals $\hat{G}(k) = 
({\rm i}\omega_{n} \hat{1} - \hat{H}(\k))^{-1}$ 
is described as, 
\begin{eqnarray}
 \label{eq:Green-function}
G_{ij}(k)=\sum_{\alpha=1}^{3} u_{i\alpha}(\k) u_{j\alpha}(\k) G_{\alpha}(k),
\end{eqnarray}
where $G_{\alpha}(k)=\frac{1}{{\rm i}\omega_{n}-
E_{\alpha}(\mbox{{\scriptsize \boldmath$k$}})}$.

 In the following, we denote the energy band described by the 
dispersion relation $E_3(\k)$ as $\gamma$-band.  
 As we have shown in Fig.~1, the $\gamma$-band crosses the Fermi level, 
and the others are below the Fermi level. 
 Therefore, the superconducting transition is induced by the 
Cooper pairing in the $\gamma$-band. 
 In this case, the \eli equation is written in terms of an  
effective interaction within the $\gamma$-band, 
\begin{eqnarray}
\lambda_{\rm e} \Delta(k) = - \sum_{k'} V(k,k') |G_{3}(k')|^{2} \Delta(k'),  
\label{eq:eliashberg-equation}
\end{eqnarray}
with 
\begin{eqnarray}
\label{eq:effective-interaction}
&& \hspace{-8mm} 
V(k,k')=\sum_{abcd} u_{a3}(\k) u_{b3}(-\k) V_{abcd}(k,k') 
u_{c3}(\k') u_{d3}(-\k'). 
\nonumber 
\\
\end{eqnarray}
 The \eli equation (eq.~(\ref{eq:eliashberg-equation})) is regarded to be 
an eigenvalue equation and $\lambda_{\rm e}$ represents the maximum 
eigenvalue.
 The superconducting transition temperature is determined by the criterion 
$\lambda_{\rm e}=1$. 

 Here, we have ignored the normal self-energy which is important for a 
quantitative estimation of \Tcf. 
 However, qualitative nature of the superconductivity, such as the 
pairing symmetry and the pairing mechanism, is not affected in 
many cases including cuprates, ruthenates and organics.~\cite{rf:yanasereview} 
 This is highly expected in case of \Cof, 
unless the electronic structure is significantly affected by the 
normal self-energy. 
 We will show that the volume of \egf, which will be denoted 
as $n_{\rm e}$ below, is an important parameter for 
the pairing symmetry. 
 Therefore, it is possible that the pairing symmetry is affected by 
the normal self-energy through the modification of $n_{\rm e}$. 
 It is, however, expected that the following results are still 
valid even in this case by regarding the $n_{\rm e}$ 
modified by the normal self-energy as a relevant parameter.

\begin{table}[htbp]
  \begin{center}
    \begin{tabular}{|c|c|c|} \hline 
A$_1$ & $s$-wave & 
$1$ 
\\\hline
E$_2$ & $d$-wave & 
$
\sin \frac{\sqrt{3}}{2}k_{\rm x} \sin \frac{1}{2}k_{\rm y}
$
\\\hline
A$_2$ & $i$-wave & 
$
\sin \frac{3\sqrt{3}}{2}k_{\rm x} \sin \frac{1}{2}k_{\rm y}
+\sin \frac{\sqrt{3}}{2}k_{\rm x} \sin \frac{5}{2}k_{\rm y}
-\sin \sqrt{3} k_{\rm x} \sin 2 k_{\rm y}
$ 
\\\hline
E$_1$ & $p$-wave & 
$\sin \frac{\sqrt{3}}{2}k_{\rm x} \cos \frac{1}{2}k_{\rm y}$
\\\hline
B$_1$ & $f_1$-wave & 
$
\sin \frac{1}{2}k_{\rm y} 
(\cos \frac{\sqrt{3}}{2}k_{\rm x} - \cos \frac{1}{2}k_{\rm y})
$
\\\hline
B$_2$ & $f_2$-wave & 
$\sin \frac{\sqrt{3}}{2}k_{\rm x} 
(\cos \frac{\sqrt{3}}{2}k_{\rm x} - \cos \frac{3}{2}k_{\rm y})
$
\\\hline
    \end{tabular}
    \caption{
Classification of the pairing symmetry in the triangular lattice. 
The first column shows the irreducible representations of D$_6$ group. 
The second column shows the notation adopted in this paper. 
The $s$-wave, $p$-wave, {\it etc} are the counterparts 
of the isotropic system. 
The third column shows the typical wave function of Cooper pairs. 
}  
  \end{center}
\end{table}

 Before showing the results, it is necessary to classify 
the pairing symmetry. 
 The symmetry of Cooper pairs is classified into $s$-, 
$p$-, $d$-wave {\it etc.} in case of an isotropic system like $^3$He. 
 For metals, the Cooper pairing is classified into the finite species 
according to the symmetry of crystals.~\cite{rf:sureview} 
 We show the classification in case of the triangular lattice in Table I. 
 We denote ``$s$-wave'', ``$d$-wave'' {\it etc.} in analogy with 
the isotropic case. 
 While the $s$-, $d$- and $i$-wave are spin singlet pairings, 
the $p$-, $f_1$- and $f_2$-wave are spin triplet pairings. 
 Note that there remains two-fold degeneracy in the $p$- and 
$d$-wave states, namely $p_{\rm x}$- and $p_{\rm y}$-wave, 
$d_{\rm xy}$- and $d_{\rm x^2-y^2}$-wave, respectively. 
 The time-reversal-symmetry-breaking is expected below \Tc 
in the $d$-wave state, as discussed in the RVB 
theory.~\cite{rf:baskaran,rf:shastry,rf:lee,rf:ogata} 
 On the contrary, time-reversal-symmetry is not necessarily broken 
in the $p$-wave case because there is an internal degree of freedom 
representing the direction of $S=1$, 
as discussed in Sr$_2$RuO$_4$.~\cite{rf:yanaseRuSO}

 The eigenvalues of the \eli equation, eq.~(\ref{eq:eliashberg-equation}) 
are classified according to the symmetry of Cooper pairs. 
 The pairing symmetry corresponding to the largest eigenvalue is 
stabilized below \Tcf. 
 Hereafter, we ignore the possibility of $s$-wave pairing because 
the strong on-site repulsion will destabilize even the extended 
$s$-wave pairing. 
 When the symmetry of crystal is lowered, some candidates in Table I 
are classified into the same irreducible representation.  
 For example, the $d_{\rm xy}$-wave and $s$-wave symmetries are included 
in the same representation for the anisotropic triangular 
lattice.~\cite{rf:Ytanakaorganic,rf:kurokiorganic} 
 However, we can ignore this possibility in the isotropic triangular 
lattice.

\subsection{Phase diagram of three-orbital model}

 In order to search possible pairing symmetries in a phase diagram, 
we introduce two controlling parameters, $a$ and $n_{\rm e}$.
 Among the hopping matrix elements in eqs.~(\ref{eq:e11}-\ref{eq:e23}), 
the largest one, namely $t_3$ is fixed to $1$ 
but the other matrix elements are chosen to be 
\begin{eqnarray}
\label{eq:minor-matrix}
&& \hspace{-10mm} (t_1,t_2,t_4,t_5,t_6,t_7,t_8,t_9)=
\nonumber \\ 
&& \hspace{-5mm} a (0.1,0.2,0.3,-0.2,-0.05,0.2,0.2,-0.25). 
\end{eqnarray}
 We choose this parameter set so that the dispersion relation 
obtained in the LDA calculation~\cite{rf:singh,rf:pickett} is 
appropriately reproduced when $a \sim 1$. 
 In case of $a=0$, the system is regarded to be a superposition of 
kagome lattice,~\cite{rf:koshibae} but we have to choose $a \geq 0.6$ 
in order to obtain a realistic Fermi surface. 
 Thus, the parameter $a$ indicates a deviation from the kagome lattice. 
 Although there are many choices of controlling the minor matrix elements,  
we have confirmed that the following results are qualitatively 
independent of the choice.

 As another controlling parameter, we use the hole number $n_{\rm e}$ 
in the $e_{\rm g}$-Fermi surface, which can be altered by adjusting 
the crystal field splitting $e_{\rm c}$. 
 When we decrease $e_{\rm c}$, the energy of $e_{\rm g}$-orbitals is 
lowered and thus $n_{\rm e}$ decreases. 
 We have confirmed that the value $n_{\rm e}$ is essential rather 
than the total electron number $n$ for the following results 
which are almost independent of the way to alter $n_{\rm e}$. 
 Note that the total electron number is fixed as $n=5.33$ 
throughout this paper.

\begin{figure}[ht]
\begin{center}
\includegraphics[width=7cm]{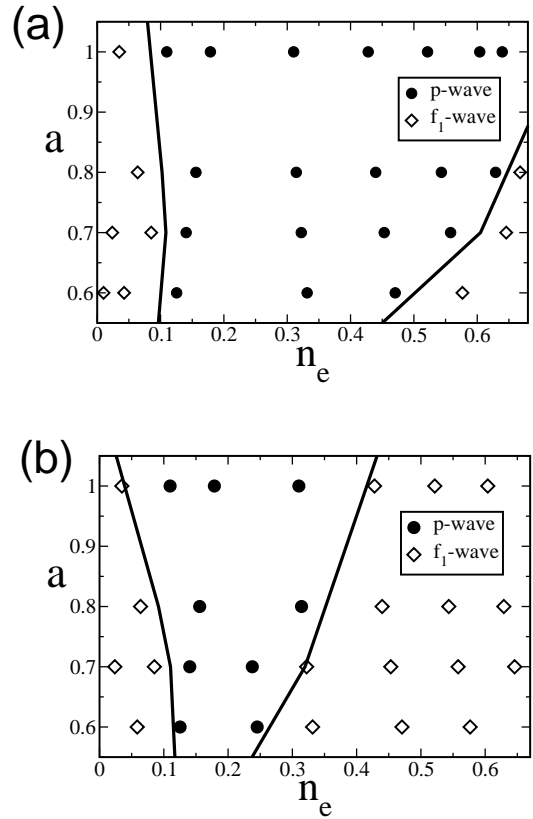}
\caption{Phase diagram for (a)$U'=\Jh=J=U/3$ and  
(b)$U'=U/2$ and $\Jh=J=U/4$. The horizontal and vertical axes are 
described in the text. 
The solid line is the phase boundary obtained by the interpolation. 
} 
\label{fig:phasediagram3D}
\end{center}
\end{figure}

 We divide the first Brillouin zone into $128 \times 128$ lattice points 
and take 512 Matsubara frequencies. 
 We have confirmed that the following results do not depend on 
the numerical details, qualitatively. 
 In the following, the temperature is fixed to be $T=0.01$ unless we 
mention explicitly. 
 It will be shown in Fig.~5 that the stable pairing symmetry is almost 
independent of the temperature. 
 We fix $U=5$ and change the value of $\Jh=J$. 
 Under the reasonable conditions $U=U'+2\Jh$ and $U'-\Jh > 0$, 
$\Jh=U/3$ is the maximum value of the Hund's rule coupling.

 Figure.~3 shows the most stable pairing symmetry in the phase diagram 
of $a$ and $n_{\rm e}$ for two values of the interaction strength. 
 As shown in Fig.~3(a), the spin triplet $p$-wave superconductivity 
is stabilized in the wide region of parameter space when $\Jh=U/3$. 
 The $f_1$-wave superconductivity is also stabilized when 
$e_{\rm g}$-Fermi surface is very small or very large. 
 For the values of $n_{\rm e}$ expected in the LDA calculation, 
namely $n_{\rm e}=0.1 \sim 0.3$, we obtain the $p$-wave 
superconductivity independent of the value of $a$. 
 When the value of Hund's rule coupling is decreased (Fig.~3(b)),  
the $f_1$-wave superconductivity becomes more stable. 
 We see that in both cases the spin triplet superconductivity is stable.

 By definition, the $e_{\rm g}$-Fermi surface vanishes in case 
of $n_{\rm e}=0$. 
 Then, it is difficult to determine the pairing state 
since the tendency to superconductivity is very weak independent of 
the pairing symmetry. 
 On the other hand, the superconductivity is not significantly 
affected by the disappearance of $a_{\rm 1g}$-Fermi surface which 
occurs at $n_{\rm e}=0.67$.

\begin{figure}[ht]
\begin{center}
\includegraphics[width=7cm]{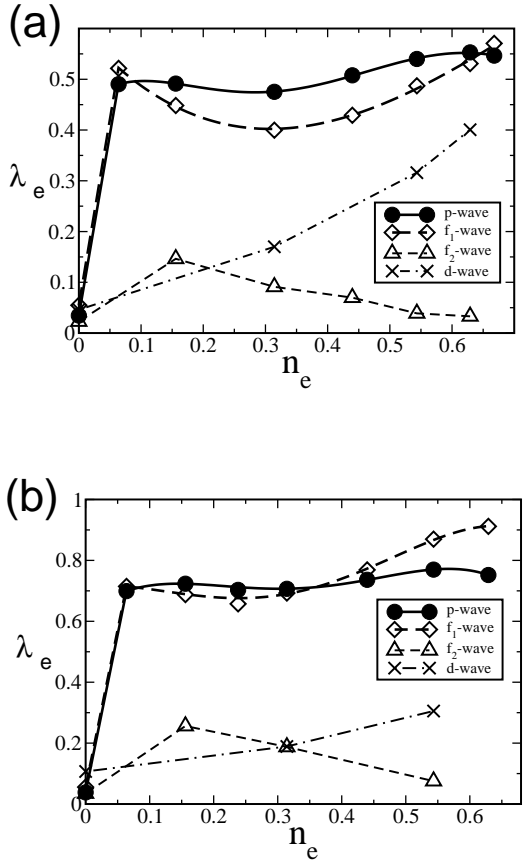}
\caption{
$n_{\rm e}$-dependence of eigenvalues of \eli equation. 
We choose $T=0.01$, $a=0.8$ and (a)$U'=\Jh=J=U/3$ or 
(b)$U'=U/2$ and $\Jh=J=U/4$.  
} 
\label{fig:ne-dependence}
\end{center}
\end{figure}

 In order to make the situation clearer, 
we show the eigenvalues of \eli equation for each pairing symmetry 
in Fig.~4. 
 It is shown that the $p$- and $f_1$-wave superconductivity have  
nearly degenerate eigenvalues in a wide parameter range. 
 If we assume the weak crystal field splitting $e_{\rm c} \sim 0$, 
we obtain $n_{\rm e} \sim 0.3$ which is consistent with LDA calculation. 
 The eigenvalue in the $f_1$-wave symmetry shows a minimum 
around this value. 
 As a result, the $p$-wave superconductivity is stable in this 
region. 
 As the Hund's rule coupling decreases, eigenvalues of 
both $p$- and $f_1$-wave symmetries increase, but that of  
the $f_1$-wave symmetry increases more rapidly (See also Fig.~7). 
 Note that the eigenvalues for the $d$-wave, $i$-wave and 
$f_2$-wave states are very small compared to the $p$- and $f_1$-wave states. 
 As is shown later, the $d$-wave state is stabilized when 
Hund's rule coupling is very small (Figs.~7 and 8).

\begin{figure}[ht]
\begin{center}
\includegraphics[width=7cm]{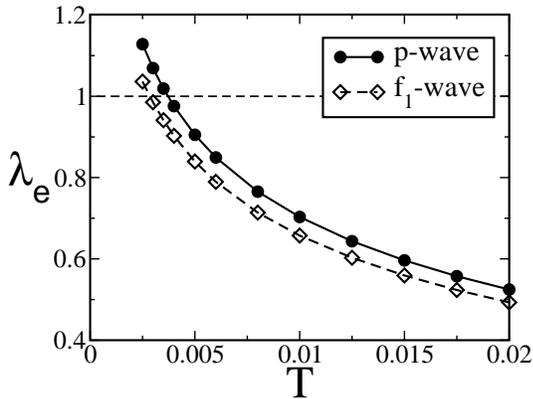}
\caption{
Temperature dependence of eigenvalues of \eli equation. 
We choose $a=0.8$, $n_{\rm e}=0.238$, $U'=U/2$ and $\Jh=J=U/4$. 
} 
\label{fig:te-dependence}
\end{center}
\end{figure}

 We see that $\lambda_{\rm e}$ is still less than $1$ at $T=0.01$ 
(Fig.~4). 
 Therefore, the pairing instability occurs at lower temperature.  
 Fig.~5 shows the temperature dependence of $\lambda_{\rm e}$ at 
$\Jh=U/4$, $a=0.8$ and $n_{\rm e}=0.238$ where the maximum eigenvalue 
is $\lambda_{\rm e} \sim 0.7$ at $T=0.01$. 
 Then, we obtain $\lambda_{\rm e}=1$ at \Tc$ = 0.0037$ 
for the $p$-wave symmetry. 
 If we assume $t_{3}=200 $meV so that the total band width 
is $W=1.8$eV, \Tc$ = 0.0037$ corresponds to \Tc$ = 8$K consistent 
with experimental value. 
 Furthermore, Fig.~5 clearly shows that most stable pairing symmetry 
is almost independent of temperature. This means that 
the phase diagram obtained at $T=0.01$ is very accurate.

 Another interesting result in Fig.~4 is that 
the maximum eigenvalue does not significantly depend on $n_{\rm e}$. 
 Even if the size of $e_{\rm g}$-Fermi surface is remarkably 
reduced, the instability of superconductivity is not suppressed 
unless the $e_{\rm g}$-Fermi surface vanishes. 
 This is mainly because the DOS of \egf little depends on the value 
of $n_{\rm e}$. 
 This is one of the characteristics of the two-dimensional system in the low 
density region. 
 Note that the number of hole included in each hole pocket is very 
small as $n_{\rm e}/6 \sim 0.05$. 
 Then, an analogy with the isotropic system like $^3$He is partly 
justified. 
 This picture is important for the pairing mechanism as we will explain
in \S3.3. 
 The $n_{\rm e}$-dependence of \Tc can be measured by 
varying the Na-content of \Cof. 
 However, experimental results seems to be 
controversial.~\cite{rf:schaak,rf:milne}

 The eigenvalue rapidly decreases when \egf vanishes.   
 This result indicates that the \egf plays an essential role 
for the superconductivity. This implication will be clearly 
confirmed in \S4.2. 
 Although the eigenvalues are very small, 
the $d$-wave symmetry seems to be most stable at $n_{\rm e}=0$. 
 Then, the topology of Fermi surface is equivalent to the 
simple triangular lattice including only the nearest neighbor hopping. 
 In this sense, our result at $n_{\rm e}=0$ is qualitatively 
consistent with the RVB theory based on the $t$-$J$ model 
in the triangular lattice, which shows the 
$d_{\rm x^{2}-y^{2}} \pm {\rm i} d_{\rm xy}$-wave 
superconductivity.~\cite{rf:baskaran,rf:shastry,rf:lee,rf:ogata} 
 However, the used parameters are quite different. The $t$-$J$ model 
assumes $U/t > 8$, while $U/t=5$ in this paper. 
 In the intermediate coupling region, the momentum dependence arising 
from the vertex correction is probably important when the SOP gives 
very small $\lambda_{\rm e}$.~\cite{rf:yanasereview}  
 In case of the simple triangular lattice, the lowest order vertex 
correction favors the $p$-wave state.~\cite{rf:nisikawa2002} 
 It should be stressed that the SOP gives much larger value of 
$\lambda_{\rm e}$ when \egf exists, as shown in Fig.~4.

\begin{figure}[ht]
\begin{center}
\includegraphics[width=7cm]{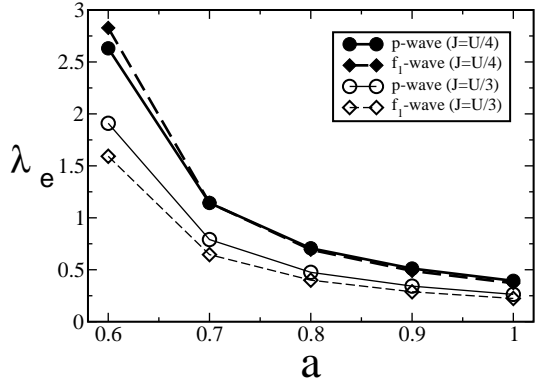}
\caption{
$a$-dependence of eigenvalues of \eli equation. 
We choose the parameter $\Jh=U/3$ or $\Jh=U/4$. 
Here, we fix the parameter $e_{\rm c}=0$ instead of $n_{\rm e}$. 
Therefore, $n_{\rm e}$ slightly differs from $n_{\rm e}=0.33$ at $a=0.6$ 
to $n_{\rm e}=0.31$ at $a=1$. 
} 
\label{fig:a-dependence}
\end{center}
\end{figure}

 Fig.~6 shows the $a$-dependence of eigenvalues. 
 It is shown that the eigenvalue monotonically increases with  
decreasing $a$. This variation is basically owing to the increase of 
the DOS. 
 In case of $a=0.5$, almost flat band is realized around the 
\egff. Therefore, a steep increase of the eigenvalue leading to the 
remarkable enhancement of \Tc occurs toward $a=0.5$. 
 We note that most important parameter for the appearance of flat band is 
the next nearest neighbor hoppings. 
 Although by changing the parameter $a$, the nearest and third nearest 
neighbor hoppings vary simultaneously, these parameters play only 
quantitative roles. 
 From Figs.~3-6, we see that the variable $a$ is important for 
the value of \Tcf, while the variable $n_{\rm e}$ plays an essential 
role for determining the pairing symmetry.

\begin{figure}[ht]
\begin{center}
\includegraphics[width=7cm]{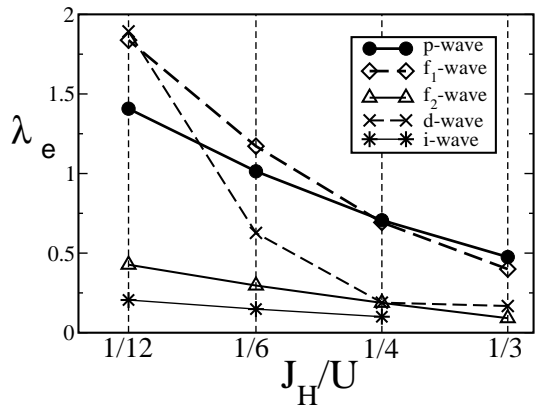}
\caption{
$\Jh$-dependence of eigenvalues of \eli equation. 
The parameters are chosen to be $a=0.8$ and $n_{\rm e}=0.31$ 
} 
\label{fig:j-dependence}
\end{center}
\end{figure}

 Before closing this subsection, let us discuss the possibility of 
$d$-wave superconductivity in case of the small Hund's 
rule coupling. 
 Fig.~7 shows the $\Jh$-dependence of eigenvalues for each pairing 
symmetry. 
 It is shown that all eigenvalues increase with the decrease 
of Hund's rule coupling. 
 Among them, the eigenvalue in the $d$-wave symmetry increases most rapidly 
and the $d$-wave superconductivity is stabilized for $\Jh < U/12$. 
 The phase diagram in the $\Jh$-$n_{\rm e}$ plane is shown in Fig.~8. 
We see that the $d$-wave superconductivity is more stable 
when $n_{\rm e}$ is small.

\begin{figure}[ht]
\begin{center}
\includegraphics[width=8cm]{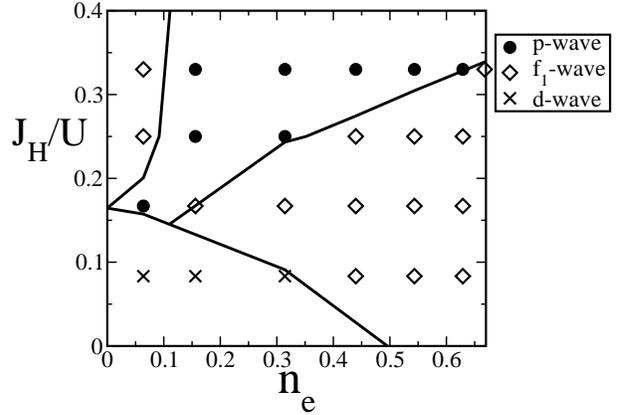}
\caption{Phase diagram in the $\Jh$-$n_{\rm e}$ plane at $a=0.8$. 
The solid line is the phase boundary obtained by the interpolation. 
} 
\label{fig:j-phase}
\end{center}
\end{figure}

 This stability of the $d$-wave pairing is basically owing to the 
large value of $U'$ which is comparable to $U$. 
 The inter-orbital repulsion $U'$ couples to the charge and orbital 
excitations which contribute to the effective interactions 
equivalently in the singlet and triplet channels. 
 Therefore, the difference between singlet and triplet superconductivity 
is reduced when $U'$ is large. 
 In other words, the Hund's rule coupling favors the spin triplet 
superconductivity, although the value of \Tc is reduced. 
 However, we expect that the $d$-wave superconductivity is less stable
if we include the higher order terms because 
higher order terms significantly enhance the spin excitation 
rather than the orbital and charge excitation. 
 In other words, the role of $U'$ will be reduced 
in the higher order theory. 
 This is confirmed by the FLEX calculation.~\cite{rf:motiduki}

\subsection{Basic mechanism of superconductivity}

 In order to clarify the basic mechanism of superconductivity, 
we study the momentum dependence of effective interaction $V(k,k')$ 
in the spin triplet channel.  
 Figure~9 shows the $\k'$-dependence of $V(k,k')$ with $\k$ being 
fixed at the momentum shown by an arrow at which the order parameter 
in the $p$-wave symmetry takes maximum value. 
 It is apparent that there is a strong attractive interaction 
between momenta included in the same hole pocket Fermi surface. 
 This is the reason why the spin triplet superconductivity is favored. 
 We can show that in case of $\Jh=U/3$, the effective interaction in 
the singlet channel has opposite sign to that in the triplet channel. 
 This strong repulsive interaction remarkably suppresses 
the spin singlet superconductivity.

\begin{figure}[ht]
\begin{center}
\includegraphics[width=8cm]{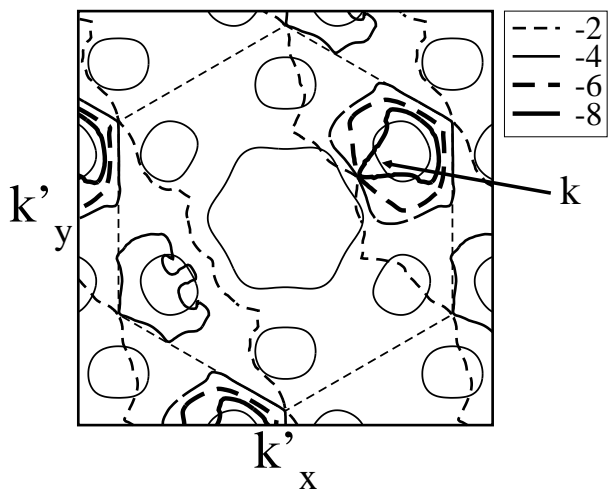}
\caption{
Contour plot of the effective interaction $V(k,k')$. 
The initial momentum $\k$ is shown in the figure. 
The horizontal and vertical axis show $k_{\rm x}'$ and $k_{\rm y}'$, 
respectively. Matsubara frequency is fixed to the lowest value 
$\omega_{\rm n}=\omega'_{\rm n}=\pi T$. 
The Fermi surface is simultaneously described by the thin solid line.  
The parameters are chosen to be $n_{\rm e}=0.31$, $a=0.8$, 
$U'=U/2$ and $\Jh=J=U/4$.  
} 
\label{fig:effectiveinteraction}
\end{center}
\end{figure}

\begin{figure}[ht]
\begin{center}
\includegraphics[width=7cm]{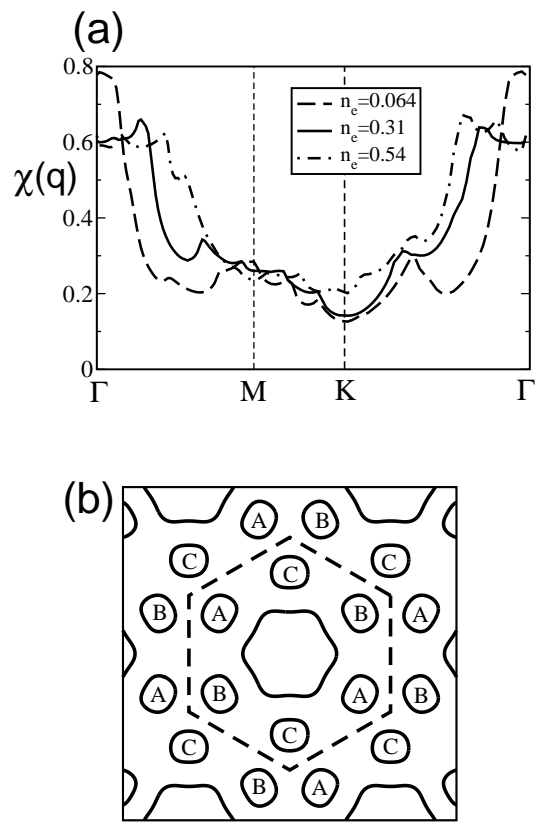}
\caption{
(a) Momentum dependence of the static spin susceptibility at $a=0.8$. 
(b) Schematic figure for the classification of hole pockets. 
} 
\label{fig:kaitotal}
\end{center}
\end{figure}

 The microscopic origin of this momentum dependence can be 
understood as follows. 
 First, we point out the ferromagnetic character of spin fluctuation. 
 Fig.~10(a) shows the spin susceptibility which is estimated by the 
Kubo formula within the bubble diagram. 
 It is clearly that the spin susceptibility has a trapezoidal peak 
around $\q=0$. 
  Note that the ferromagnetic spin fluctuation has been expected 
in the LDA calculation~\cite{rf:singh} and observed by the 
NMR measurement~\cite{rf:ishida}. 
 Owing to the ferromagnetic character of spin susceptibility, 
the attractive interaction in the same hole pocket is very strong and 
favors the spin triplet superconductivity.

 The ferromagnetic spin fluctuation is basically comes from the 
\egff. 
 Each hole pocket gives rise to the ferromagnetic spin fluctuation 
like in the two-dimensional electron gas, which has a susceptibility 
with the trapezoidal structure. 
 Actually, as shown in Fig.~10(a), when we increase the size of hole 
pockets by changing $n_{\rm e}$, the width of the trapezoidal peak 
around the $\Gamma$ point increases.

 Next, we illuminate the essential roles of the orbital degree of 
freedom. 
 First, we point out that the ferromagnetic spin fluctuation is
indeed induced by the orbital degree of freedom. 
 In the multi-orbital model, the spin susceptibility is determined 
by the dispersion relation and the structure factor arising from 
the orbital degree of freedom. 
 If we neglect the momentum dependence of structure factor as was 
done in the previous studies,~\cite{rf:johannes,rf:nisikawa} 
we obtain two peaks of spin susceptibility which are quit 
different from ours. 
 One is located around the M point and the other is slightly removed 
from the $\Gamma$ point. 
 However, we obtain the trapezoidal peak centered at the $\Gamma$ point 
by appropriately taking account of the structure factor. 
 Thus, the frustration inherent in the triangular lattice is 
removed by the orbital degree of freedom which gives rise to 
the ferromagnetic spin fluctuation.

 Second, we point out that the roles of the orbital degree of freedom 
can be understood by considering the momentum dependence of 
the wave function which is expressed by the unitary matrix 
$\hat{U}(\k)$ in eq.~(\ref{eq:unitary}). 
 This wave function indicates the orbital character of quasi-particles 
(see also \S4). 
 The structure factor of spin discussed above is also obtained by 
this wave function. 
 Furthermore, the effective interaction $V(k,k')$ has another distinct 
property arising from this momentum dependence. 
 As we have mentioned before, the \egf mainly consists of 
the $e_{\rm g}$-doublet whose wave function is shown in 
eqs.~(\ref{eq:e1}) and (\ref{eq:e2}). 
 Furthermore, we find that the six hole pockets are divided 
into three pairs as is shown in Fig.~10(b). 
 For example, more than $90$\% of the weight of wave function 
in the Fermi surface ``A'' originates from the orbital 
$|e_{\rm g}, 1>$, while the other two pairs are 
dominated by respective linear combinations of 
$|e_{\rm g}, 1>$ and $|e_{\rm g}, 2>$. 
 It is generally expected that the electron correlation between 
the same orbitals is stronger than that between the different orbitals. 
 Actually, the effective interaction between different pairs 
``A'', ``B'' and ``C'' is significantly smaller than those between 
the same pairs,  as shown in Fig.~9. 
 This is the reason why the $p$- and $f_1$-wave superconductivities 
are stabilized with nearly degenerate eigenvalues as shown in Fig.~4. 
 Which is more stable between $p$- and $f_1$-wave states depends on the 
coupling between different pairs of hole pockets, which is generally 
small as explained above. 
 Note that if we apply the phenomenological theory on 
the ferromagnetic spin-fluctuation-induced superconductivity to \Cof, 
the $f_1$-wave superconductivity is much more stable rather than 
the $p$-wave superconductivity. 
 The single band model leading to the ferromagnetic spin 
fluctuation~\cite{rf:kuroki} also concludes the $f_1$-wave symmetry. 
 However, the $p$-wave superconductivity can be stabilized 
in the present case owing to the orbital degeneracy.

 It should be noticed that the origin of trapezoidal peak of spin 
susceptibility around $\Gamma$ point is clearly understood 
by this momentum dependence of wave function. 
 Although the wave functions are not orthogonal between different pairs 
of hole pockets, the matrix elements between them in calculating 
$\chi(q)$ are small. Therefore, in the zeroth order 
approximation, pairs of hole pockets are regarded to be decoupled from 
each other. 
 Then, each hole pocket independently induces 
the trapezoidal peak of $\chi(q)$ as in the two-dimensional 
electron gas model.

 Another point to stabilize the superconductivity is the disconnectivity 
of the \egf as discussed before the discovery of 
\Cof.~\cite{rf:kuroki2001} 
 Even in the anisotropic superconductivity such as 
$p$-wave or $f_1$-wave symmetry, the order parameter can 
take a same sign in each hole pocket, 
which stabilizes the superconductivity induced by the ferromagnetic 
spin fluctuation. 
 Note that the difficulty of the ferromagnetic 
spin-fluctuation-induced superconductivity (superfluidity) 
has been discussed for $^3$He.~\cite{rf:2dparamagnon} 
 This difficulty is removed by the topological aspect of Fermi surface 
in case of \Cof.

\subsection{Momentum dependence of superconducting gap}

 Next, we show the momentum dependence of 
order parameter $\Delta(\k,{\rm i}\pi T)$ in Fig.~11. 
 Although $\lambda_{\rm e}$ does not reach $1$ at $T=0.01$ (Fig.~5), 
it is generally expected that the amplitude 
$|\Delta(\k,{\rm i}\pi T)|$ shows the momentum dependence of 
superconducting gap below \Tc and determines the low energy excitation.  
 We note that even if the superconducting instability is dominated 
by the \egff, the \agf also contributes to the low energy excitations 
observed by NMR $1/T_1T$, specific heat and magnetic field penetration 
depth.

 Fig.~11(a) shows the order parameter in the $p$-wave symmetry. 
 We choose the Hund's rule coupling as $\Jh=U/3$ where the $p$-wave 
superconductivity is stabilized. 
 Among the two degenerate $p_{\rm x}$- and $p_{\rm y}$-states, 
only the $p_{\rm y}$-state is shown. 
 Because of the discontinuity of the \egff, the order parameter is 
node-less on the \egff, while it has nodes on the \agff. 
 Since $p_x \hat{x} \pm p_y \hat{y}$, $p_x \hat{y} \pm p_y \hat{x}$ 
or $(p_x \pm {\rm i} p_y) \hat{z}$ states are expected below \Tcf, 
the superconducting gap becomes 
$\sqrt{\Delta_{\rm x}(k)^{2}+\Delta_{\rm y}(k)^{2}}$, 
where $\Delta_{\rm x}(k)$ and $\Delta_{\rm y}(k)$ are the order 
parameters for $p_{\rm x}$- and $p_{\rm y}$-states, respectively. 
 In this case, the superconducting gap does not vanish even on the 
\agff. But, we find a remarkable anisotropy of the superconducting gap 
on the \agf which can explain the power-law behaviors of NMR $1/T_{1}T$ 
and so on, like in the case of Sr$_2$RuO$_4$.~\cite{rf:nomuragap} 
 However, we note that this is an accidental result.

 Fig.~11(b) shows the order parameter in the $f_1$-wave symmetry. 
 We choose the Hund's rule coupling as $\Jh=U/6$ where the $f_1$-wave 
superconductivity is most stable. 
 We can see the clear six times alternation of the sign of order 
parameter. Also in this case, the \egf is node-less and \agf has  
line nodes. 
 As we showed before for the magnetic penetration depth,~\cite{rf:uemura}
the combination of fully gaped \egf and line nodes on \agf gives an 
intermediate temperature dependence between $s$-wave and anisotropic 
superconductivity.

\begin{figure}[ht]
\begin{center}
\includegraphics[width=7cm]{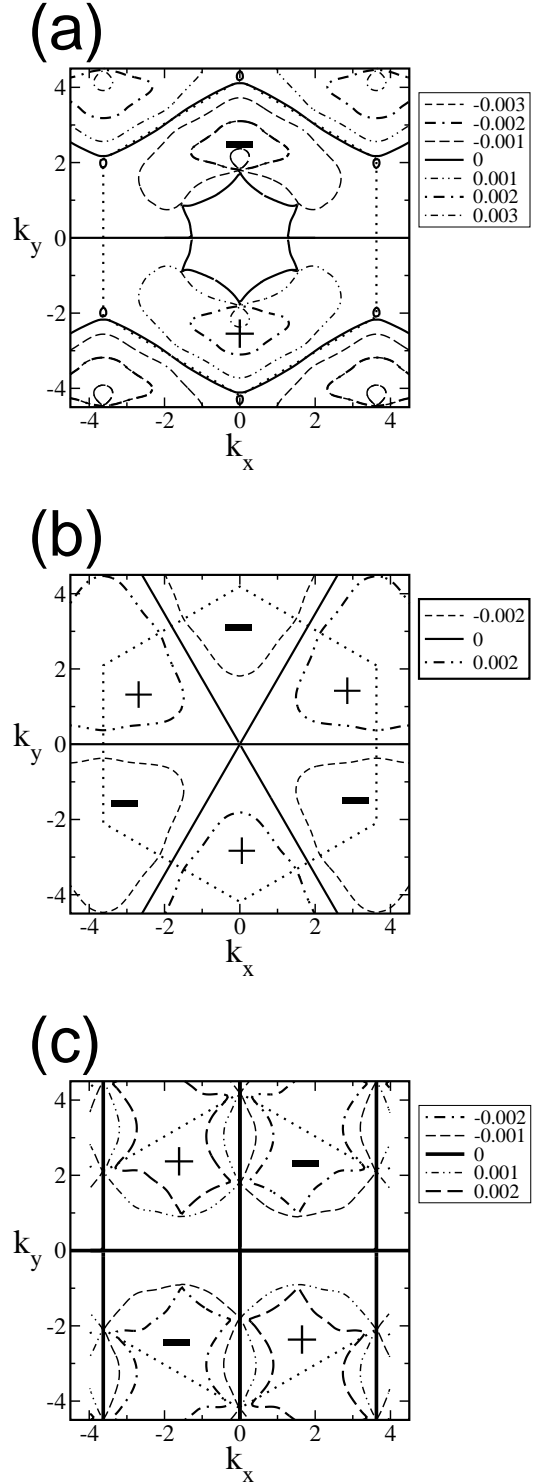}
\caption{Momentum dependence of order parameter (a) in the 
$p_{\rm y}$-wave symmetry, (b) in the $f_1$-wave symmetry and 
(c) in the $d_{\rm xy}$-wave symmetry. 
The parameters are chosen to be $a=0.8$ and $n_{\rm e}=0.31$. 
} 
\label{fig:wavefunction}
\end{center}
\end{figure}

 In Fig.~11(c) we show the order parameter in the $d_{\rm xy}$-wave
state which is stabilized when $\Jh$ is very small, $\Jh=U/12$.  
 The $d_{\rm xy} \pm {\rm i}d_{\rm x^{2}-y^{2}}$ state is 
expected below \Tc and both \agf and \egf are node-less in this case. 
 The exponential behaviors in many quantities are expected unless some 
accidental situation occurs as in the $p$-wave state. 
 Our calculation does not support such an accidental situation 
in the $d$-wave symmetry.

 It should be noticed that in all of the above cases we have shown, 
the amplitude of order parameter is large on the \egff, 
while it is small on the \agff. 
 This result is expected from the fact that the \egf is responsible for 
the pairing instability as discussed in \S3.3. 
 This point will be illuminated more clearly in the next section.

\section{Reduced Models}

  We have analyzed the possibility of unconventional 
superconductivity in \Co on the basis of the three-orbital model. 
 Because calculations for this model need much computational time, 
a simplified model appropriate for studying the superconductivity is 
highly desired for a future development in the theoretical field. 
 In this section, we try to find an appropriate model 
from the comparison to the three-orbital model. 
 We show that the two-orbital model is satisfactory for this purpose,  
while the single-orbital model is not. 
 The essential origin of the results in \S3 will be clarified by 
these trials.

\subsection{Failure of single-orbital Hubbard model}

 Thus far, we have stressed some essential roles of the orbital 
degeneracy. They are illuminated by showing the failure of 
single-orbital model. 
 Some authors have already studied single-orbital Hubbard models 
reproducing the LDA Fermi surface.~\cite{rf:nisikawa,rf:kuroki} 
 In this paper, we try a single-orbital Hubbard model by keeping 
only the $\gamma$-band, {\it i.e,} the highest-energy eigenstates 
obtained in eq.~(\ref{eq:unitary}).  
Hamiltonian is expressed in the following way. 
\begin{eqnarray}
H_{1} =\sum_{\k,s} E_{3}(\k) c_{\k,s}^{\dag} c_{\k,s} + 
U \sum_{i} n_{i,\uparrow} n_{i,\downarrow}. 
\label{eq:single-orbital-model}
\end{eqnarray}
 As has been shown in Fig.~1, the typical Fermi surface is reproduced 
in this model. 
 Indeed, this is the minimal model describing the electron correlation 
in this material. 
 However, as shown below, this model is inappropriate for the study of 
superconductivity because the results are qualitatively 
different from those in the multi-orbital model.

 We clarify the term ``single-orbital Hubbard model'' in order to avoid 
any confusion. 
 In this paper, ``single-orbital Hubbard model'' suggests 
the single-band model including only the {\it momentum independent} 
interaction like eq.~(\ref{eq:single-orbital-model}). 
 As is shown later, we can construct a single-band model in which the 
roles of orbital degeneracy are appropriately represented in the 
momentum dependence of interaction term. 
 Thus, we distinguish ``single-orbital Hubbard model'' from 
`single-band model''.

\begin{figure}[ht]
\begin{center}
\includegraphics[width=7cm]{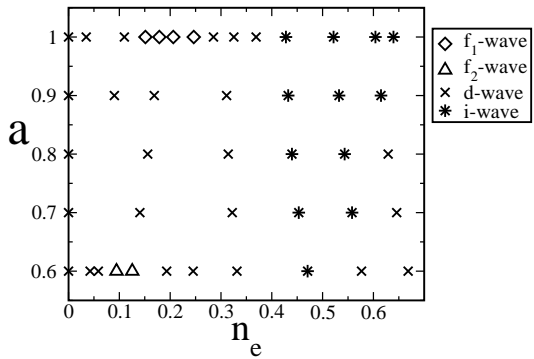}
\caption{Phase diagram of the single-orbital Hubbard model. 
The qualitatively different results from Fig.~3 indicate the failure of 
this model. 
} 
\label{fig:singlebandmodel}
\end{center}
\end{figure}

 In Fig.~12, we show the phase diagram obtained by the SOP applied to the 
single-orbital Hubbard model (eq.~(\ref{eq:single-orbital-model})). 
 In the wide region of parameter space, the $d$-wave and $i$-wave 
superconductivities are stabilized instead of 
$p$-wave and $f_1$-wave states. 
 The $f_1$-wave superconductivity competes with the $d$-wave one, 
but is stabilized only in a narrow region. 
 The $p$-wave superconductivity is not stabilized 
in the whole parameter range.

 This difference arises from the disregard of the momentum dependence
of wave function which is represented by $\hat{U}(\k)$. 
 If we neglect the momentum dependence of $\hat{U}(\k)$ 
in eq.~(\ref{eq:unitary}), the three-orbital model is reduced 
to the single-orbital Hubbard model in eq.~(\ref{eq:single-orbital-model}). 
 The difference of stable pairing state is apparent 
if we check the spin susceptibility $\chi(q)$. 
 In the single-orbital Hubbard model, 
$\chi(q)$ is similar to that obtained in Ref.~31 and 
we do not clearly see the ferromagnetic tendency (see also the
discussion in \S3.3).  
 As a result, the momentum dependence of the effective interaction 
is qualitatively different form that in the three-orbital model.

 This difference is partly improved by neglecting the 
$a_{\rm 1g}$-orbital like Ref.~30. 
 Then, we obtain the nearly ferromagnetic spin fluctuation and 
spin triplet superconductivity. 
 However, the coupling between different pairs of hole pocket Fermi 
surfaces (see Fig.~10(b)) is over-estimated, and therefore, 
the $f_1$-wave state is stabilized much more than the $p$-wave state. 
 This is not consistent with the results in \S3. 
 We wish to stress again that the characteristic nature of orbital 
in each hole pocket Fermi surface induces the nearly degeneracy 
between the $p$-wave and $f_1$-wave states. 
 This characteristic nature can not be taken into account in the 
single-orbital Hubbard model.

\subsection{Effective two-orbital model}

 The results in the previous subsection show that 
the single-orbital Hubbard model is qualitatively inappropriate 
for studying the superconductivity. 
 The important factor to be taken into account is the orbital character 
of quasi-particles on each Fermi surface. 
 This is described by the momentum dependence of the unitary matrix 
$\hat{U}(\k)$ in eq.~(\ref{eq:unitary}). 
 Considering these points, we propose a simplification of 
the three-orbital model in this subsection. 
 The reduced model is an effective two-orbital model representing the 
$e_{\rm g}$-doublet. 
 The simplification is performed by the following two steps. 

(i) The $a_{\rm 1g}$-orbital is simply ignored. 

(ii) The lower band below the Fermi level is ignored. 
\\
 The first step is justified because we find that the superconducting 
instability is dominated by the six hole pocket Fermi surfaces 
which mainly  consist of the $e_{\rm g}$-orbitals. 
 The second one is generally justified because the quasi-particles 
around the Fermi surface lead to the superconductivity.

 In order to perform the first step, we transform the basis of 
local orbitals. This is carried out by using the unitary 
transformation as, 
\begin{eqnarray}
 \label{eq:unitary-local}
&& \hspace{-20mm} 
(d_{\k,1,s}^{\dag},d_{\k,2,s}^{\dag},d_{\k,3,s}^{\dag})=
(c_{\k,1,s}^{\dag},c_{\k,2,s}^{\dag},c_{\k,3,s}^{\dag}) \hat{U}_{\rm l}, 
\\
\hspace{-30mm} 
&& \hat{U}_{\rm l} =
\left(
\begin{array}{ccc}
\frac{1}{\sqrt{3}} & 0 & \frac{2}{\sqrt{6}} \\
\frac{1}{\sqrt{3}} & \frac{1}{\sqrt{2}} & -\frac{1}{\sqrt{6}} \\
\frac{1}{\sqrt{3}} & -\frac{1}{\sqrt{2}} & -\frac{1}{\sqrt{6}} \\
\end{array}
\right). 
\end{eqnarray}
 The interaction term $H_{\rm I}$ in the Hamiltonian $H_{3}$ is 
invariant for this unitary transformation owing to the relations 
$U=U'+2 \Jh$ and $\Jh=J$. 
 The non-interacting term is transformed as, 
\begin{eqnarray}
 \label{eq:transformed-non-interactiong-part}
&& \hspace{-10mm} 
H_0 = \sum_{\k,s} d_{\k,s}^{\dag} \hat{H}'(\k) d_{\k,s}, 
\\
&& \hspace{-10mm} 
\hat{H}'(\k) = \hat{U}_{\rm l}^{\dag} \hat{H}(\k) \hat{U}_{\rm l}.
\end{eqnarray}
 The first step is performed by dropping the creation (annihilation)
operator $d_{\k,1,s}^{\dag}$ ($d_{\k,1,s}$) which corresponds to the 
$a_{\rm 1g}$-orbital. 
 As a result, the three-orbital model is reduced to the 
following two-orbital model. 
\begin{eqnarray}
&& \hspace{-10mm}
H_{2} = \sum_{\k,s} a_{\k,s}^{\dag} \hat{h}(\k) a_{\k,s} 
+ U \sum_{i} \sum_{a=1}^{2} n_{i,a,\uparrow} n_{i,a,\downarrow} 
 \nonumber \\ 
&& \hspace{-5mm}
+ U' \sum_{i} \sum_{a>b} n_{i,a} n_{i,b} 
- \Jh \sum_{i} \sum_{a>b} (2 \SS_{i,a} \SS_{i,b} + \frac{1}{2} n_{i,a} n_{i,b})
 \nonumber \\ 
&& \hspace{-5mm}
+ J \sum_{i} \sum_{a \neq b}  
a_{i,a,\downarrow}^{\dag} 
a_{i,a,\uparrow}^{\dag} 
a_{i,b,\uparrow} 
a_{i,b,\downarrow}. 
\label{eq:two-orbital-model}
\end{eqnarray}
 Here, we have introduced a $2 \times 2$ matrix 
$\hat{h}(\k)_{i,j}=\hat{H}'(\k)_{i+1,j+1}$ and 
two component vector 
$a_{\k,s}^{\dag}=(d_{\k,2,s}^{\dag},d_{\k,3,s}^{\dag})$. 
 Then, the Green function is described by a $2 \times 2$ matrix as 
$\hat{G}(k)=({\rm i}\omega_{n} \hat{1} - \hat{h}(\k))^{-1}$, 
whose elements are expressed as 
\begin{eqnarray}
 \label{eq:2by2-Green-function}
G_{ij}(k)=\sum_{\alpha=1}^{2} v_{i\alpha}(\k) v_{j\alpha}(\k) G_{\alpha}(k). 
\end{eqnarray}
 Here, $v_{i\alpha}(\k)$ are components of the unitary matrix 
$\hat{V}^{\dag}(\k)$ which diagonalizes the matrix $\hat{h}(\k)$
\begin{eqnarray}
 \label{eq:unitary2by2}
\hat{V}^{\dag}(\k)  \hat{h}(\k) \hat{V}(\k) 
= 
\left(
\begin{array}{cc}
e_1(\k) & 0 \\
0 & e_2(\k) \\
\end{array}
\right), 
\end{eqnarray}
with $e_1(\k)<e_2(\k)$. 
 The diagonalized Green function is obtained as 
$G_{\alpha}(k)=\frac{1}{{\rm i}\omega_{n}-
e_{\alpha}(\mbox{{\scriptsize \boldmath$k$}})}$. 

 We show the dispersion relation $e_1(\k)$ and $e_2(\k)$ in Fig.~13. 
 Apparently the band structure around the \egf is unchanged 
by this simplification, while the \agf vanishes.

\begin{figure}[ht]
\begin{center}
\includegraphics[width=7cm]{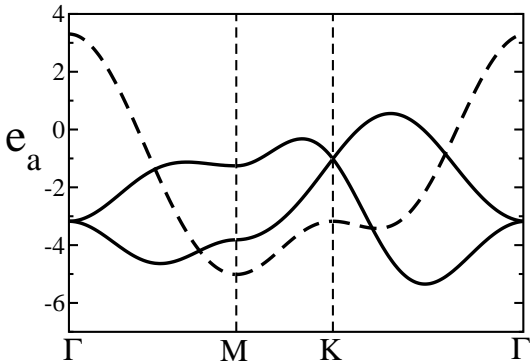}
\caption{
Dispersion relation in the two-orbital model (solid lines). 
The parameters are chosen to be $a=0.8$ and $n_{\rm e}=0.36$. 
We have shown the dispersion relation of the $a_{\rm 1g}$-orbital 
which is obtained as $\hat{H}'(\k)_{11}-\mu$ (dashed line). 
} 
\label{fig:fermisurface2by2}
\end{center}
\end{figure}

 The second step is performed by ignoring the lower energy band, 
$e_1(\k)$. 
 Then, the Green function is obtained as, 
$G_{ij}(k)=v_{i2}(\k) v_{j2}(\k) G_{2}(k)$. 
 Owing to this procedure, the calculation becomes equivalent to that for 
a single band Hamiltonian with momentum-dependent interaction, 
\begin{eqnarray}
&& \hspace{-10mm}
H_{L} =\sum_{\k,s} e_{2}(\k) c_{\k,s}^{\dag} c_{\k,s} 
\nonumber \\
&& \hspace{-5mm}
+ \sum_{\q,\k',\k} S(\q,\k',\k) 
             c_{\q-\k,\uparrow}^{\dag} c_{\q-\k',\downarrow}^{\dag} 
             c_{\k',\downarrow} c_{\k,\uparrow}
\nonumber \\
&& \hspace{-5mm} 
+ \sum_{\q,\k',\k,\sigma} S'(\q,\k',\k) 
             c_{\q-\k,\sigma}^{\dag} c_{\q-\k',\sigma}^{\dag} 
             c_{\k',\sigma} c_{\k,\sigma}. 
\label{eq:long-range-model}
\end{eqnarray}
 The momentum dependent factors $S(\q,\k',\k)$ and $S'(\q,\k',\k)$ are 
expressed by the Coulomb interactions $U$, $U'$, $\Jh$ and $J$ and 
the wave function $v_{i2}(\k)$. 
 If we neglect the momentum dependence of unitary matrix 
$\hat{V}^{\dag}(\k)$, the factor $S(\q,\k',\k)$ becomes $U$ and 
$S'(\q,\k',\k)=0$.  
 Then, the model is exactly reduced to the single-orbital Hubbard model 
described by eq.~(\ref{eq:single-orbital-model}) with use of 
$e_{2}(\k)$ instead of $E_{3}(\k)$. 
 We have discussed in \S4.1 that this single-orbital 
Hubbard model is not appropriate. 
 On the other hand, the Hamiltonian $H_{L}$ is appropriate because 
the roles of orbital degeneracy are taken into account in 
the momentum dependence of interaction.

\begin{figure}[ht]
\begin{center}
\includegraphics[width=7cm]{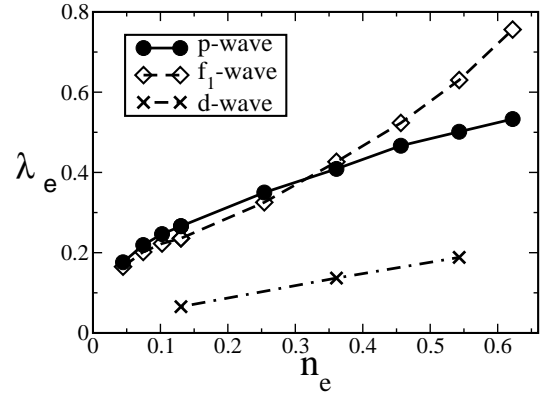}
\caption{Eigenvalues of \eli equation obtained 
in the effective two-orbital model. 
We choose the parameters $a=0.8$ and $\Jh=U/4$. 
} 
\label{fig:phasediagram2by2}
\end{center}
\end{figure}

 We find that the results for the superconductivity are almost the same 
between the Hamiltonian $H_{2}$ and $H_{L}$. 
 In Fig.~14 we show the $n_{\rm e}$-dependence of eigenvalues of \eli 
equation for the simplified model, $H_{L}$. 
 We see that the increase of eigenvalues with $n_{\rm e}$ is steeper 
than that in Fig.~4. 
 This is mainly owing to the increase of DOS. 
 However, the relation between each pairing symmetry closely resembles. 
 For example, the $p$-wave superconductivity is stable around 
$n_{\rm e}=0.2$, while the $f_1$-wave superconductivity is realized for 
larger values of $n_{\rm e}$. 
 The eigenvalue for the spin singlet $d$-wave superconductivity is 
far below that for the spin triplet one. 
 These results mean that the effective two-orbital model described by 
eq.~(\ref{eq:two-orbital-model}) or eq.~(\ref{eq:long-range-model}) 
appropriately reproduces the results in the three-orbital model. 
 The fact that the step (1) is appropriate clearly means that the 
superconductivity is basically led by the \egff.  
 The \agf plays only a secondary role.

 Note that the eigenvalue of \eli equation decreases owing to the step (1), 
mainly because the DOS in the \egf decreases. 
 We have confirmed that the step (2) slightly enhances the 
spin triplet superconductivity.

\section{Effects of Vertex Corrections in a Two-Orbital Model}

 In this section, we study the effects of vertex corrections. 
 Although it is desirable to study these effects in the three-orbital 
model, we use the effective two-orbital model whose validity has been 
demonstrated in \S4.2, because of numerical difficulties. 
 Generally speaking, the higher order terms may play an 
important role for the superconducting instability, 
since it is considered that most of unconventional superconductors 
are in the intermediate coupling region. 
 For example, vertex correction which is not included in the RPA 
plays an important role to stabilize the spin triplet pairing in 
Sr$_2$RuO$_4$.~\cite{rf:nomura} 
 Therefore, it is an important issue to investigate the role of 
higher order corrections in the present model.

\begin{figure}[ht]
\begin{center}
\includegraphics[width=8cm]{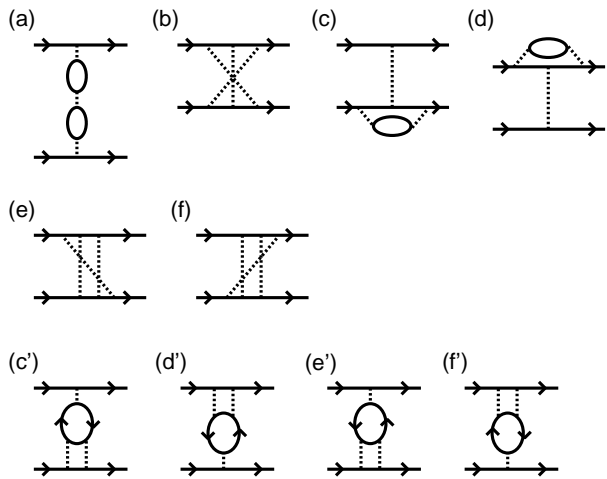}
\caption{Diagrammatic representation of the third order terms in the 
effective interaction. 
(a-f) correspond to the spin singlet channel or spin triplet channel 
with d-vector $d \parallel {\hat z}$. 
(c'-f') correspond to the spin triplet channel with d-vector 
$d \perp {\hat z}$. 
} 
\label{fig:3rddiagram}
\end{center}
\end{figure}

 We apply the third order perturbation theory (TOP) and 
its renormalized version to the Hamiltonian $H_2$ 
(eq.~(\ref{eq:two-orbital-model})). 
 We adopt this model instead of more simplified model $H_{\rm L}$ 
(eq.~(\ref{eq:long-range-model})) because the computational time 
is hardly reduced by the second step (ii) in \S4.2. 
 The parameter is chosen to be $\Jh=U/3$, where the interaction between 
electrons with same spin vanishes and thus the number of diagrams is 
much reduced. 
 As discussed in \S3.2, this region will be relevant rather than 
the region where the Hund's rule coupling is small.

 Fig.~15 shows the diagrammatic representation of third order terms 
in the effective interaction. 
 Figs.~15(a) and (b) are classified into the RPA terms and others are 
the vertex corrections. 
 The present theory is invariant for the rotation of spin, 
since we do not take account of the spin-orbit interaction. 
 Therefore, the result on the spin triplet pairing does not depend on 
the direction of $d$-vector. 
 Note that two RPA terms cancel each other in case of the 
spin triplet pairing with $d \parallel z$.

\begin{figure}[ht]
\begin{center}
\includegraphics[width=7cm]{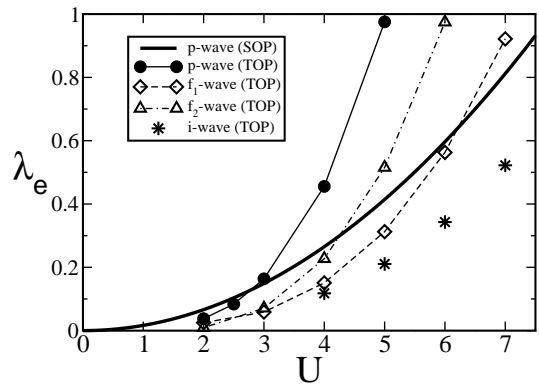}
\caption{
Eigenvalues of \eli equation in the third order perturbation theory. 
The thick solid line shows the maximum eigenvalue in the second order 
perturbation theory, which is classified into the $p$-wave symmetry. 
We do not show the eigenvalue in the $d$-wave symmetry because 
the tendency to superconductivity is very weak. 
We fix the parameters $a=0.6$, $n_{\rm e}=0.35$ and $\Jh=U/3$. 
} 
\label{fig:naivetop}
\end{center}
\end{figure}

 We numerically solve the \eli equation within the TOP and show the 
eigenvalues in Fig.~16. 
 We see that the $p$- and $f_2$-wave superconductivity are significantly 
stabilized for $U>4$, while the $f_1$-wave and spin singlet pairings are 
unfavored. 
 However, as discussed below, we find that these results in the 
intermediate coupling region are fictitious. 
 Within the third order terms in Fig.~15, dominant contributions 
for triplet channel come from the terms represented 
in Figs.~15(e') and (f'), which include a particle-particle ladder. 
 In contrast, the terms represented in Figs.~15(c') and (d') with a 
particle-hole ladder are negligible. 
 As is well known in the Kanamori theory on the metallic 
ferromagnetism,~\cite{rf:kanamori} the particle-particle ladder diagrams 
generally induce the screening of interaction as 
$U \rightarrow U(q)=U/(1+U\phi(q))$ where $\phi(q)$ is
obtained by the particle-particle ladder diagram.  
 If $q$-dependence of $U(q)$ is not important, this scattering process is 
incorporated by the renormalized coupling constant $\bar{U}$. 
 In the above TOP calculation, only the lowest order term in the 
Kanamori-type correction was taken into account. 
 Therefore, it is reasonable to think that the contributions from 
Figs.~15(e') and (f') can be suppressed if we include the higher order 
perturbation terms.

\begin{figure}[ht]
\begin{center}
\includegraphics[width=6cm]{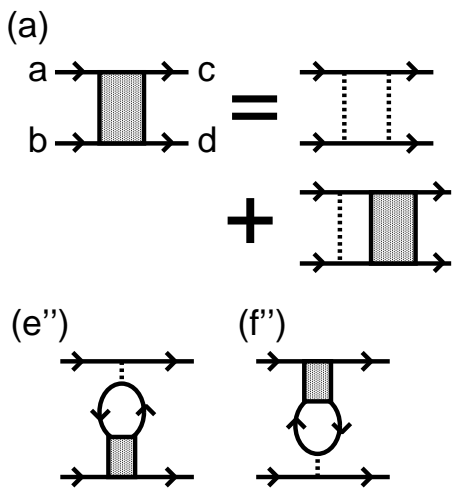}
\caption{
(a) Renormalization of the particle-particle ladder diagram. 
In (e'') and (f''), the renormalized particle-particle ladder is used 
instead of bare ladder. 
In the RTOP, we take account of the terms (e'') and (f'') instead of 
(e') and (f') in Fig.~15. 
} 
\label{fig:t-matrix}
\end{center}
\end{figure}

 In order to investigate this possibility, we perform a calculation 
of a renormalized TOP (RTOP), as shown in Fig.~17. 
 The particle-particle ladder in Figs.~15(e') and (f') are 
replaced by the T-matrix shown in Fig.~17(a). 
 As a result, the infinite order terms representing the screening effect 
are taken into account as in the Kanamori theory. 
 By using the diagrams in Figs.~2, 15(c',d') and 17(e'',f''), 
we estimate the effective interaction and solve the \eli equation. 
 The obtained eigenvalues are shown in Fig.~18. 
 It is apparent that the results of naive TOP is significantly altered 
by the renormalization and that the correction to the SOP is small. 
 In particular, the $p$-wave superconductivity is slightly stable over 
the $f_1$-wave superconductivity. 
 The nearly degeneracy between these states is also reproduced. 
 The order parameter in each pairing symmetry is very similar to Fig.~11, 
although that in the naive TOP is remarkably different. 
 We see that the eigenvalues are slightly reduced from the SOP, however 
the $U$-dependence is almost unchanged. 
 These results are naturally interpreted if we consider that the 
vertex corrections basically work as a screening effect. 
 Then, the second order perturbation theory is justified 
by regarding the interactions to be the renormalized ones.

\begin{figure}[ht]
\begin{center}
\includegraphics[width=7cm]{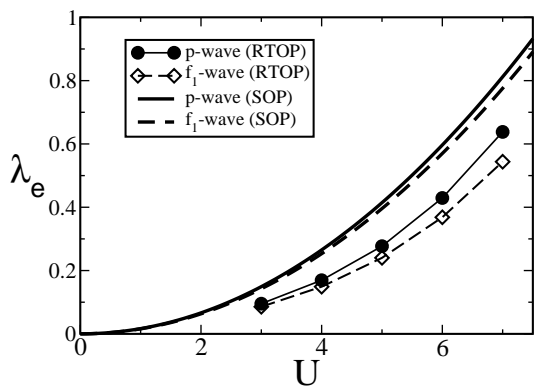}
\caption{
Eigenvalues of \eli equation in the renormalized third order 
perturbation theory for $p$-wave (circles) and $f_1$-wave (diamonds) 
symmetry. 
Note that the eigenvalue in the $f_2$-wave symmetry is very small. 
The thick solid and dashed lines show the eigenvalues in the 
second order perturbation theory for the $p$-wave and $f_1$-wave 
symmetry, respectively. The parameters are the same as in Fig.~16. 
} 
\label{fig:rtop}
\end{center}
\end{figure}

 Let us compare the present results to the case of high-\Tc cuprates and 
Sr$_2$RuO$_4$. 
 For high-\Tc cuprates, the $d$-wave superconductivity is basically 
induced by the RPA terms and the vertex correction due to 
the particle-particle ladder diagrams effectively reduces 
the coupling constant.~\cite{rf:yanasereview,rf:bulut} 
 Therefore, the situation is very similar to the present case, 
although there is a difference of singlet and triplet pairing. 
 On the other hand, in case of Sr$_2$RuO$_4$, the effective interaction 
derived from the RPA terms has very weak momentum dependence, 
which does not work for the anisotropic pairing. 
 However, the $q$-dependence of particle-particle ladder in TOP favors 
the spin triplet superconductivity.~\cite{rf:nomura} 
 Then, the naive discussion on the screening effect can not be applied. 
 It has been confirmed that the qualitative results of TOP applied 
to Sr$_2$RuO$_4$ are not altered even when the renormalization of 
particle-particle ladder is taken into account.~\cite{rf:nomuraforth} 
 Thus, the basic mechanism of possible spin triplet 
superconductivity in \Co is qualitatively different from 
that in Sr$_2$RuO$_4$.

\section{Discussions}

 In this paper, we have investigated the multi-orbital model 
for \Co on the basis of the perturbation theory. 
 The obtained results indicate a possibility of 
spin triplet superconductivity in this material, 
although the $d$-wave superconductivity is also stabilized in a part of 
parameter space. 
 There are two candidates of spin triplet pairing; 
$p$-wave and $f$-wave superconductivity are nearly degenerate.

 Although the spin triplet superconductivity is one of the most 
interesting issues in the condensed matter physics, the microscopic 
theory remains in the developing stage. 
 This is mainly owing to very few $d$-electron materials showing the 
spin triplet superconductivity. 
 Although we see many candidates in the heavy fermion materials, 
the theoretical treatment is generally difficult for $f$-electron 
systems. 
 Therefore, a discovery of spin triplet superconductor in 
transition metal oxides will lead to an important development 
in the microscopic understandings. 

 Probably, most established spin triplet superconductor 
in $d$-electron systems is Sr$_2$RuO$_4$.~\cite{rf:maeno} 
 Therefore, we have provided detailed discussions on the comparison 
between Sr$_2$RuO$_4$ and \Cof. 
 According to the results in this paper, \Co provides a qualitatively 
different example from Sr$_2$RuO$_4$ in the following two points. 

 First, the RPA terms give rise to the dominant scattering process 
leading to the spin triplet pairing. 
 The spin excitation is clearly ferromagnetic and favorable for the 
spin triplet pairing. 
 This is in sharp contrast to the case of Sr$_2$RuO$_4$ where the vertex 
corrections are essential for the $p$-wave pairing. 
 In case of \Cof, the vertex corrections induce only the screening 
effect which is not important for the qualitative results. 
 While the ferromagnetic spin-fluctuation-induced spin triplet 
superconductivity has been discussed from early years, the 
corresponding superconductivity has not been established until now. 
 We expect that \Co will be a first example realizing this mechanism.

 Second, the orbital degeneracy plays an essential role 
in case of \Cof. 
 The conduction band in \Co as well as that in Sr$_2$RuO$_4$ 
are basically described by three t$_{\rm 2g}$-orbitals. 
 Although the single-orbital Hubbard model is an appropriate model 
for describing the pairing mechanism of Sr$_2$RuO$_4$,~\cite{rf:nomura} 
such a simplification is qualitatively inappropriate for \Cof. 
 The success of single-orbital Hubbard model for Sr$_2$RuO$_4$ is due to 
the electronic structure where the $\gamma$-band is basically described 
by the local $d_{\rm xy}$-orbital. 
 The failure for \Co is due to the fact that the \egf can not 
be described by any individual local orbital. 
 In other words, the hybridization term in the unperturbed Hamiltonian 
is large in case of \Cof, while it is negligible in Sr$_2$RuO$_4$ 
owing to the particular crystal symmetry. 
 In this sense, \Co will be a more typical example of 
the multi-orbital superconductor.  
 Then, the momentum dependence of the wave function of quasi-particles 
essentially affects the effective interaction 
leading to the Cooper pairing.

 We have pointed out that the reduced two-orbital model is appropriate, 
instead of the failure of single-orbital model. 
 This is because the Fermi surface in \Co can be classified according 
to the local orbitals. 
 Then, the superconductivity is basically triggered by the \egff . 
 Since a portion of $a_{\rm 1g}$-orbital in the \egf is less than 5\%, 
this orbital is safely ignored. 
 This situation is similar to the case of Sr$_2$RuO$_4$. 
 However, the orbital degeneracy in $e_{\rm g}$-doublet 
can not be ignored in case of \Cof.

 From the above comparisons, we obtain the following empirical rules.  

(1) When the RPA-terms are favorable for the anisotropic superconductivity, 
the non-RPA terms are not qualitatively important, and {\it vice versa}. 

(2) When a part of Fermi surface is described by a few local orbitals, 
the simplification of microscopic model is possible. 

 In particular, the second rule will be helpful for a future 
development of microscopic understanding on the 
multi-band superconductors. 
 For example, several Fermi surfaces appear in heavy fermion materials. 
 This fact as well as the 14-fold degeneracy in $f$-shell make the 
microscopic treatment difficult. 
 However, it will be possible to obtain a simplified model by  
identifying the microscopic character of each Fermi surface.

 Thus far, we have discussed the superconductivity in \Co induced 
by the electron-electron correlation and highlighted 
the possibility of spin triplet pairing.  
 However, any clear experimental evidence for the symmetry of 
superconductivity has not been obtained up to now. 
 Instead, we see some experimental observations which restrict 
the pairing state. 
 For example, the absence of (or very small) coherence peak in 
NMR $1/T_{1}T$,~\cite{rf:yoshimura,rf:kobayashi,rf:ishida,rf:zheng} 
power-law temperature dependence of 
$1/T_{1}T$~\cite{rf:ishida,rf:zheng} 
and specific heat,~\cite{rf:hdyang,rf:lorenz,rf:oeschler}
NMR Knight shift below \Tc ~\cite{rf:yoshimura,rf:kobayashi,
rf:ishidaprivate,rf:zhengprivate} and time-reversal symmetry observed 
in $\mu$SR~\cite{rf:higemoto} should be cited, although a part of 
them are controversial. 
 As for the results in this paper, 
spin triplet $p$- or $f_1$-wave superconductivity is consistent with 
the absence of coherence peak and with the power-law behaviors 
below \Tc. 
 In both cases, the (quasi-)line nodes appear in the \agff. 
 In case of the $p$-wave pairing, the time-reversal-symmetry observed 
in $\mu$SR indicates a $d$-vector parallel to the plane, namely 
$\hat{d}=p_{\rm x}\hat{x} \pm p_{\rm y}\hat{y}$ or 
$\hat{d}=p_{\rm x}\hat{y} \pm p_{\rm y}\hat{x}$. 
 This direction of $d$-vector is consistent with the recent measurements 
of NMR Knight shift under the parallel 
field~\cite{rf:kobayashi,rf:ishidaprivate,rf:zhengprivate} 
as well as macroscopic $H_{\rm c2}$,~\cite{rf:chou,rf:sasaki} 
if we assume that the $d$-vector is strongly fixed 
against the applied magnetic field. 
 We note that the qualitatively different result has been obtained 
in the NMR Knight shift,~\cite{rf:yoshimura} which is consistent with 
this pairing state if the $d$-vector is weakly fixed 
against the magnetic field.

 Although we have shown that the $d$-vector in Sr$_2$RuO$_4$ 
is very weakly fixed against the magnetic field,~\cite{rf:yanaseRuSO} 
this is partly owing to the particular electronic structure of 
Sr$_2$RuO$_4$. Therefore, we expect that the anisotropy of $d$-vector 
is larger for \Cof. 
 The symmetry breaking interaction leading to the anisotropy arises 
from the second order term with respect to the spin-orbit interaction 
for Sr$_2$RuO$_4$, while it arises from the first order term 
in case of \Cof. Therefore, it is possible that the $d$-vector 
is strongly fixed against the magnetic field in case of \Cof. 
 Quantitative estimations for the anisotropy will be one of 
the interesting future issues.

 On the other hand, the possibility of spin singlet superconductivity 
has not been denied up to now. 
 Then, the absence of time-reversal symmetry breaking will be a 
issue to be resolved for $d$-wave pairing because the 
$d_{\rm x^{2}-y^{2}} \pm {\rm i} d_{\rm xy}$ state is expected so as 
to gain the condensation energy. 
 The local distortion of triangular lattice or the feedback effect 
will be a candidate of the resolution. 
 It seems that the $i$-wave superconductivity~\cite{rf:kurokiprivate} 
is consistent with the present experimental results except for the 
very weak impurity effects.~\cite{rf:yokoi} 
 However, the microscopic mechanism leading to the pairing 
with $T_{\rm c}=5$K will be difficult for such a high 
angular momentum state. In our study, we have not found the stable 
$i$-wave state. 
 Although the observed impurity effect seems to support the 
$s$-wave pairing which is robust for the disorder, very short 
quasi-particle life time or significant anisotropy in the gap function 
has to be assumed for the absence of coherence peak in $1/T_{1}T$. 
 We consider that further vigorous investigations are highly 
desired for the identification of pairing state in \Cof.

\section*{Acknowledgments}

 The authors are grateful to Y. Ihara, K. Ishida, M. Kato, Y. Kitaoka, 
K. Kuroki, Y. Kobayashi, C. Michioka, M. Sato, Y. Tanaka, Y. J. Uemura 
and G-q. Zheng for fruitful discussions. 
 Numerical computation in this work was partly carried out 
at the Yukawa Institute Computer Facility. 
 The present work was partly supported by a Grant-In-Aid for Scientific 
Research from the Ministry of Education, Science, Sports and Culture, Japan.

\end{document}